\documentstyle[12pt]{article}    
\begin{document}
\begin{center}
Vortex Clustering: The Origin of the Second Peak in the \\ 
Magnetisation Loops of Type Two Superconductors \\
\vspace{1cm}
\noindent{[Short title: Vortex Clustering]}\\
\vspace{1cm}

D.K. Jackson$^{a}$, M. Nicodemi$^{a}$, G. K. Perkins$^{b}$, 
N.A.Lindop$^{c}$, Henrik Jeldtoft Jensen$^{a}$
\footnote{Corresponding author, email: h.jensen@ic.ac.uk}
\vspace{1cm}

$^a$ Department of Mathematics, Queen's gate 180, SW7 2BZ, United Kingdom,\\
$^b$ Blackett Laboratory, Imperial College, Prince Consort Road,
SW7 2BZ, United Kingdom, \\
$^c$ Department of Chemistry, University of Edinburgh, Edinburgh EH9 3JJ, 
United Kingdom\\
\end{center} 


\begin{center} Abstract \end{center}
We study vortex clustering in type II Superconductors. We demonstrate
that the ``second peak'' observed in magnetisation loops may be a
dynamical effect associated 
with a density driven instability of the vortex system. 
At the microscopic level the instability shows up 
as the clustering of individual vortices at (rare) preferential regions of the pinning potential. In the limit of quasi-static ramping the instability
is related to a phase transition in the equilibrium vortex system.

\newpage

When the external magnetic field penetrates  a type II superconductor, 
magnetic vortex lines appear inside the bulk of the sample. These vortex lines
 are repulsive and quantised  each carrying a multiple $q$ of the magnetic flux
 quantum $\phi_0 $. The energy per unit length of a vortex line is 
proportional to $q^2$ and accordingly  one typically expects to have 
configurations with well separated single quantised vortex  lines 
\cite{tinkham}. We show below, however, that  dynamical vortex 
clustering may  become very  important: vortices form inhomogeneous spatial 
structures which become relevant for the evolution of the system. 
This clustering may for instance
cause the ``second peak'' observed 
in magnetisation loops  and strongly  affects the structure of the 
vortex system, magnetic relaxation,
 and the distribution of the local magnetic induction as measured 
in $\mu$ spin-relaxation experiments.
The understanding of structural properties of vortices, underlying 
the presence of the ``second peak'', is one of the central issues of 
current research in superconductivity, and is related to fundamental aspects 
of vortex matter ranging from dynamical behaviour to phase transitions.
(see eg. \cite{second-peak-static,second-peak-dynamics}).  

In a superconducting sample in the presence 
of an external magnetic field, vortex 
lines penetrating from the surface into the bulk may be
trapped on pinning centres \cite{tinkham,note1}
leading to a spatially 
inhomogeneous vortex distribution  and to a net magnetisation of the sample.
As the external field is increased the vortex lines are squeezed together
and therefore interact more strongly. Such a strong interaction 
will counteract the  pinning forces and thus one expects, 
as in fact is usually observed, the magnetisation  to decrease with increasing 
external field once vortices have fully penetrated the sample.

However, upon further increase of the external field the  magnetisation is 
often observed to increase again \cite{cohen}.  This behaviour leads to the 
so-called fishtail structure in the magnetisation  data for YBCO or the 
arrowhead structure in equivalent data for BISCCO  samples\cite{Werner,1?}.
A similar peak structure has also been 
observed in low temperature superconductors \cite{Esquinazi}.
 The second peak  is one of  the most  important unsolved problems in vortex physics.  It has been seen as a signature of a phase transition in the vortex system  and its explanation has often been attempted by focusing on collective  aspects of pinning \cite{cohen,second-peak-static} or relaxation effects related to the sweeping of the field\cite{second-peak-dynamics} .  
Here we present evidence that dynamical effects are of crucial importance for
the second peak, nevertheless the mechanism inducing the peak is, in the quasi-static limit, related to a density driven phase transition in the vortex system.

In Fig. 1 we show for reference an example of half magnetisation loops
in YBCO  for a set of different temperatures. One sees how the maximum of the second peak moves to higher magnetic fields as the temperature is  decreased. 
This suggests that the mechanism  behind the increase in the magnetisation 
occurring after the first peak must also be active at zero-temperature. 
We demonstrate below that the origin of the second peak may be related to the possible  grouping of vortices  at favourable regions in the random pinning 
potential, leading to large local fluctuations in the vortex density.
  This ``clustering''
can occur in superconductors for which the ratio, $\kappa=\lambda/\xi$, between the magnetic penetration depth $\lambda$ and the coherence length $\xi$ is not too small (from our Molecular Dynamics simulations we expect roughly $\kappa>10$). 
The clustering is a dynamical effect  {\em induced by the ramping} of  the external field, as is done in magnetisation experiments.
As vortices entering (or leaving) the sample approach another vortex trapped at a position in the pinning potential three possibilities can occur. 
Firstly, the trapped vortex may be pushed ahead if the vortex-vortex repulsion
is strong enough. Secondly, the approaching vortex may move
around the trapped vortex. Or, if the trapped vortex is pinned by a force stronger than the maximum vortex-vortex repulsion, the approaching vortex may
move into the favourable position in the pinning landscape in the immediate
vicinity of the already trapped vortex.  In this way the clustering may 
significantly enhance
the effect of the rare strong pinning regions in an otherwise weak pinning background.
 
For simplicity we here consider straight parallel flux lines at separation 
$r$ for which the interaction energy per unit length is \cite{brandt1}  
\begin{equation}
U_{vv}(r)= {\phi^2_0\over 2\pi\lambda'^2}
\left[ K_0(r/\lambda')- K_0(r/\xi')\right] \ ,
\label{energy}
\end{equation}
where $\lambda ' = \lambda/\sqrt{1-b}$ and $\xi ' = \xi/\sqrt{2(1-b)}$ are the 
effective field dependent  London penetration depth and coherence length, 
respectively. $b=B/B_{c_2}$ is the reduced magnetic induction relative to 
the upper critical induction $B_{c_2}$, and  $K_0$ is a modified Bessel 
function.  Vortices can be brought close together because of the  attractive second term representing the interaction between the vortex cores.
The maximum repulsive force calculated from Eq.(\ref{energy}) is an 
upper limit for the repulsive  force between two vortex lines.    
The relative tilting and  wiggling of vortex lines will  lead to a significant decrease in the repulsion (see Eg. \cite{brandt1}).  Therefore, we expect
the clustering effect to be at least as significant in three dimensions as we demonstrate here the effect to be in two dimensions.  
In fact clustering of vortices has been directly  observed in electron microscopic imaging by Tonomura \cite{2?}.

In Fig. 2  we show the magnetisation obtained from zero temperature
Molecular Dynamics (MD) 
simulations. The vortex interaction is given by Eq.(\ref{energy}) and it is 
cut off at half the system size. We use over damped dynamics in a square two dimensional system, with periodic 
boundary conditions, of sides $100 \xi$.   The external field is ramped by 
introducing vortices into a central strip with no pins \cite{nori}; the 
magnetisation plot is calculated by considering the average density of 
vortices and its gradient in the pinned region.  
Fig. 2 clearly shows that an upturn in the magnetisation 
occurs as the external field is increased 
above the penetration field. The increase in the magnetisation coincides with 
the appearance of vortices clustered within areas of order $\xi^2$. It is  
important to mention that if the second term in Eq. \ref{energy} is left out\cite{note}, (see dashed line in Fig. 2.), clustering of vortices cannot occur and no significant upturn in the magnetisation is observed in the
simulations. At the highest field densities the potential in Eq.(\ref{energy}) 
loses its validity. This regime is also difficult to
handle numerically. Hence we do not study the full half loop in the MD 
simulations but use below a simplified lattice model to study
the increasing as well as the decreasing leg of the magnetisation. 

Let us now describe how the clustering of vortices prevents the magnetisation,  
$M$, from decaying with increasing field. We recall that approximately 
$M\propto j_c$, where  $j_c$ is the critical current density produced by the 
volume pinning force $F_p=B j_c$.  When a pinning centre becomes occupied by 
one or more trapped vortices the attractive short range pinning centre is effectively transformed into a longer range repulsive centre.
 Other vortices approaching 
this pin will feel a repulsive vortex-vortex force proportional to the number 
of trapped vortices at a distance $\lambda$  rather than the attractive force 
of range $\xi$ from the initially ``empty'' pinning centre. The local pinning 
strength will fluctuate through the sample with typically a high density of weak pinning centres and only a few sparse local strong regions. Vortices clustered 
at the few strong pinning regions can then form spatially extended energy 
barriers which cage other diffusing vortices. We emphasise that this picture is 
somewhat schematic and that in reality dynamical and collective effects are 
important. Vortices are moving in an ever changing energy landscape produced by 
the combined effect of the static spatial pinning potential and the 
instantaneous metastable configuration of the interacting vortices. The 
importance of dynamical effects follows from the fact that the detailed form, 
especially the width, of the magnetisation loops dependence on the ramping rate 
of the external field. This is the case in experiments (see  ref.s 
in \cite{cohen,second-peak-static,bhatta}) as well as in our
 simulations \cite{Mario}.

To study in detail the change of the effective energetic panorama  due to 
clustering and its consequences for the ``second peak'' we consider now a 
schematic model. This approach is 
close in spirit to similar lattice systems introduced to describe fluxons in 
superconductors, see e.g. \cite{lat_mod}. In particular we study an extension of a {\em coarse grained} cellular-automaton-like model recently introduced by  
Bassler and Paczuski (BP) \cite{Bassler}. We consider a  simplified version of a many body system with pair interactions given in eq.~(\ref{energy}) representing
a lattice  model of repulsive particles in a pinning potential  and  in contact
with a particle reservoir at a given density. Since our model explicitly allows multiple occupancy of lattice sites up to a value $N_{c2}$, we call it a 
Restricted Occupancy  Model (ROM). We apply Monte Carlo dynamics and use a Hamiltonian of the form:
\begin{equation}
{\cal H}= \frac{1}{2} \sum_{ij} n_i A_{ij} n_j 
-\frac{1}{2} \sum_i A_{ii} n_i - \sum_i A^p_i n_i
\label{H}
\end{equation}
Here $n_i\in\{0,...,N_{c2}\}$ is an integer occupancy variable equal to the 
number of particles on site $i$. The parameter $N_{c2}$ plays the role
of $B_{c2}$, it bounds the particle density per site below a critical value 
(here $N_{c2}=27$). The first term in eq.(\ref{H}) represents the  repulsive 
vortex interaction energy. The second term in eq.(\ref{H}) just normalises the particle self-interaction energy. Here we for simplicity choose the
coarse graining length to be of order the zero temperature London penetration length $\lambda(0)$.
This allows us to relate the restriction number $N_{c2}$ to the upper 
critical field $B_{c2}$ in the following way $N_{c2}=B_{c2}\lambda(0)^2/\phi_0$
where $\phi_0=hc/2e$ is the magnetic flux quantum. 
With this choice of coarse graining length it is natural as a first
approximation to assume: 
$A_{ii}=A_v=1$, $A_{ij}=A_n$ if $i$ and $j$ are nearest neighbours and 
$A_{ij}=0$ for all other couples of sites. 
We will below briefly discuss the validity of this approximation of 
the $A_{ij}$ matrix.
The third term in eq.(\ref{H}) represents a 
random pinning potential acting on a fraction $\rho=0.5$ of the lattice 
sites with $A^p=0.5$  and $A^p=0$ elsewhere.  
(The same set of interactions is used in the BP model \cite{Bassler}).
Two opposite sides of our square system ($L=32^2$) \cite{note_2} 
are in contact with a reservoir at a given  density $N_{ext}$. Particles are
introduced and  escape the system through the reservoir only. 
 
Fig. 3 shows the results of our Monte Carlo simulations of this model. We 
ramp $N_{ext}$ and record the magnetisation, $M=N_{in}-N_{ext}$ 
(with $N_{in}=\langle\sum_i n_i\rangle/L^2$), as a function of $N_{ext}$. 
The ramping of the external reservoir density, $N_{ext}$, is simply done 
by increasing it from zero up to some given value (and then decreasing back) 
by a sequence of small increments, $\Delta N_0$. 
After each increment the system  is let to relax for a time $\tau$ (in unit of 
Monte Carlo sweeps). This corresponds to a sweep rate of the applied field of 
$\gamma=\Delta N_0/\tau$. 

We recorded magnetisation loops for several values of the ratio 
$\kappa^*\equiv(\ln A_v/A_n)^{-1}$ which controls the interaction potential. 
The parameter $\kappa^*$ is qualitatively similar to $\kappa=\lambda/\xi$. 
When $\kappa^*$ is large enough a definite second peak appears in $M$. These 
magnetisation loops looks qualitatively similar to the experimental loops, 
though the peak position and amplitude of peaks in Fig. 3 are much more 
asymmetric than the peaks of the experimental data in Fig. 1. Nevertheless, 
typically magnetisation loops in other experimental samples and compounds are 
clearly asymmetric, see e.g. \cite{cohen,bhatta}.  In fact even Fig. 1 
exhibits, on careful inspection, slight asymmetries. 

The detailed loop shape depends
not only on the choice of $\kappa^*$, as shown in Fig. 3, but also on the
value of $N_{c2}$ and the ramping rate, $\gamma$. 
We find that in the $\gamma\rightarrow0$ limit, the second peak location, 
is associated with a sharp jump in 
$M_{eq}\equiv\lim_{\gamma\rightarrow 0} M(\gamma)$, corresponding to 
a true transition \cite{Mario}. The precise nature of this transition is 
currently under study. The transition occurs above the melting transition 
and leads to a significant increase in the effective energy barriers 
experienced by the diffusing vortices.

Increasing $N_{c2}$ (corresponding to higher values of $B_{c2}$) increases the 
separation between the first and the second peak. The specific features
of the loops do
also depend on the ratio between the characteristic relaxation time of the vortex system and the ramping rate,
for details see \cite{Mario,Mario2}. Quantitative differences between
the simulations and experiments are to be expected. One reason is that the
interaction strength between vortex lines depends on the magnetic induction
and the temperature. As for instance in the London approximation of Eq.(\ref{energy}) through the field and temperature dependence of $\lambda '$ and $\xi '$. This corresponds to a field and temperature dependence of the individual elements of the coupling 
matrix $A_{ij}$ in Eq.(\ref{H}). Moreover, the effective vortex screening 
length, $\lambda '$ (see Eq. \ref{energy}), increases with increasing magnetic 
field. This effect implies that non-zero $A_{ij}$ elements between 
sites of separation larger than nearest neighbour may become relevant 
as the field is increased. Interestingly, however, the present simple
approximation captures the qualitative features of the magnetic properties. 

We emphasise that according to the above picture the second peak is a dynamical effect associated with the vortices being forced in and out of the 
sample. It is linked to a density driven instability of the vortex system. 
At the more microscopic level of  description used in the Molecular Dynamics
 simulations the instability shows up 
as the clustering of individual vortices at (rare) preferential regions of the pinning potential. In the coarse grained description of the considered lattice model the instability is related to an underlying density driven phase 
transition of the equilibrium system. The instability induces
 a dramatic change of the effective collective 
energy  landscape encountered by the diffusing vortices. This, in turn, 
enormously enhances equilibrium times \cite{Mario,Mario2} and induces the 
presence of significant more spatial disorder, in strict correspondence with 
glass formers \cite{angell}.

This clustering or density instability can produce the second peak in the 
magnetisation loops. It can be thought of as a type of strong plastic 
deformation, an effect which should be observable in neutron scattering or 
$\mu$-spin resonance experiments  probing the distribution of local magnetic induction. 

DKJ and NL are grateful to acknowledge EPSRC and Imperial 
College Centre for High Temperature Superconductivity for  studentships. HJJ 
is supported by EPSRC.

\newpage

\begin{center} Figures  Captions \end{center}

\noindent {\bf  Fig. 1 } 
An example of magnetisation data and the fishtail effect 
in a single crystal of YBa2Cu4O8 at
temperautres 20K 50K and 65K. The magnetic moment is measured continuously
using a vibrating sample magnetometer while the applied field is ramped
from zero to eight Tesla and back to zero at a rate 10mT/sec . The 124
system is neither  susceptible to twinning or oxygen inhomogeneity (unlike
YBa2Cu3O7-d) indicating that the fishtail effect is intrinsic to the
pinning of point-like disorder in the crystal lattice.

\noindent {\bf  Fig. 2}
Magnetisation ($dB/dx$) vs field ($B$) for simulation with $\kappa$ of
$67$. The external field is ramped by adding one vortex between
relaxation intervals of $40$ where each time step is a maximum of
$0.01$. Fields calculated for $\xi$ of $15$ \AA. 
Solid line represents the system interacting through Eq. 1
(soft core). The dashed line represents the system interacting 
through Eq. 1 with last term omitted (hard core). The short-long dashed
curve represents the proportion of stacked vortices in the soft core case.   

Pinning centres are represented
by Gaussian wells of width $\xi$ and of amplitude $0.3 \xi^2$ 
times the condensation energy.
In this simulation they can exert a maximum pinning force 
(at zero external field) of
$4\cdot 10^{10}A/cm^2$ for the $\kappa=10$ and $8\cdot 10^8A/cm^2$ for
$\kappa=100$.   

\noindent {\bf  Fig. 3 } The magnetisation, $M$, as a function of the applied 
field density, $N_{ext}$, in the 2D R.O.M. model for  
$\kappa^*= 0.43, 0.76, 0.79$. 
The ramp rate for $N_{ext}$ is $\Delta N_0/\tau =10^{-3}$.


\begin{thebibliography}{99}
\bibitem{note1} We are here neglecting the contribution from the reversible  
magnetisation.
\bibitem{note} The second term in Eq.(\ref{energy}), 
has hitherto been left out in most studies,
see e.g. \cite{nori}.
\bibitem{note_2} We have also considered a triangular lattice to check that 
our general results do not depend on the particular lattice properties.
\end{thebibliography}
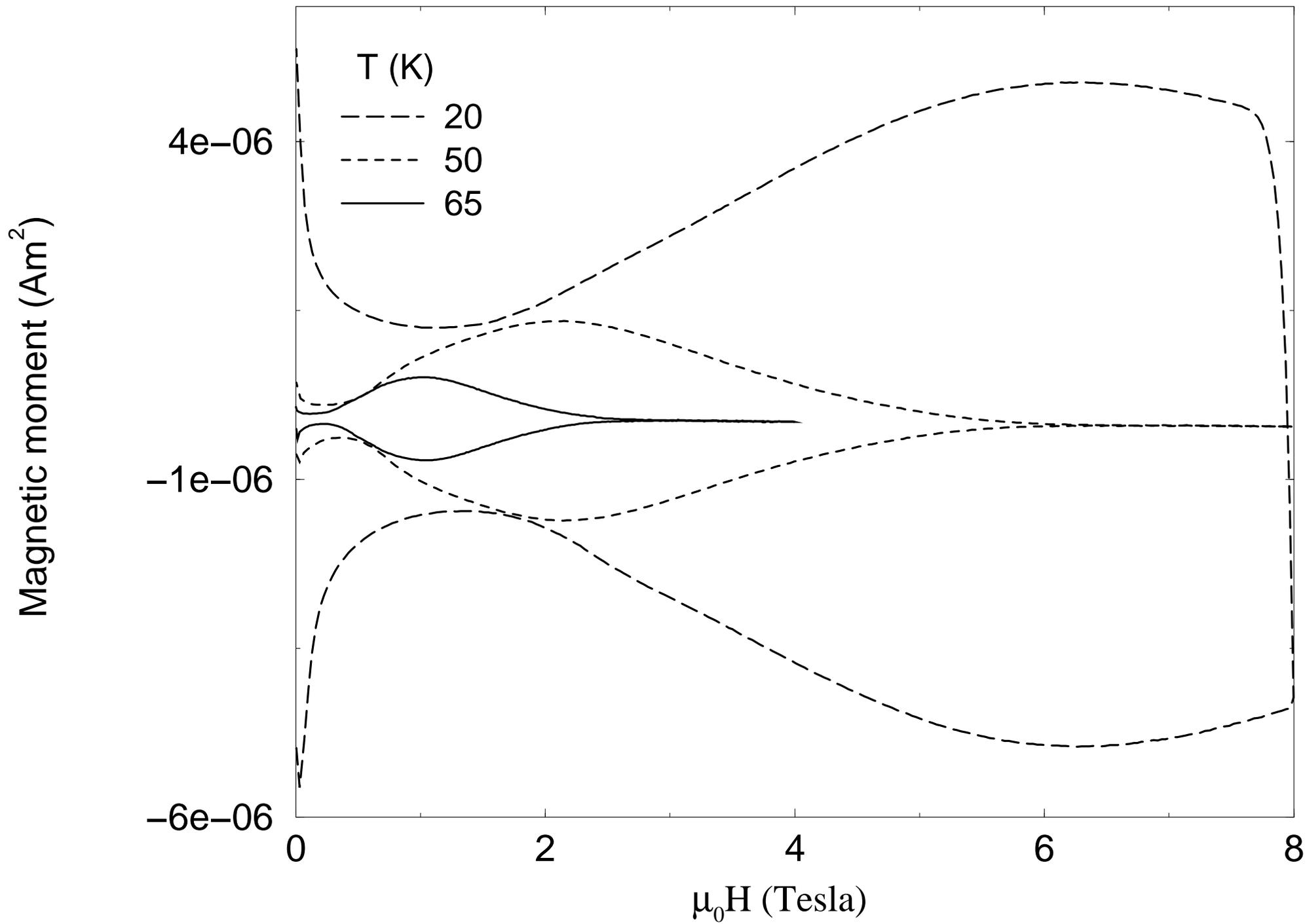
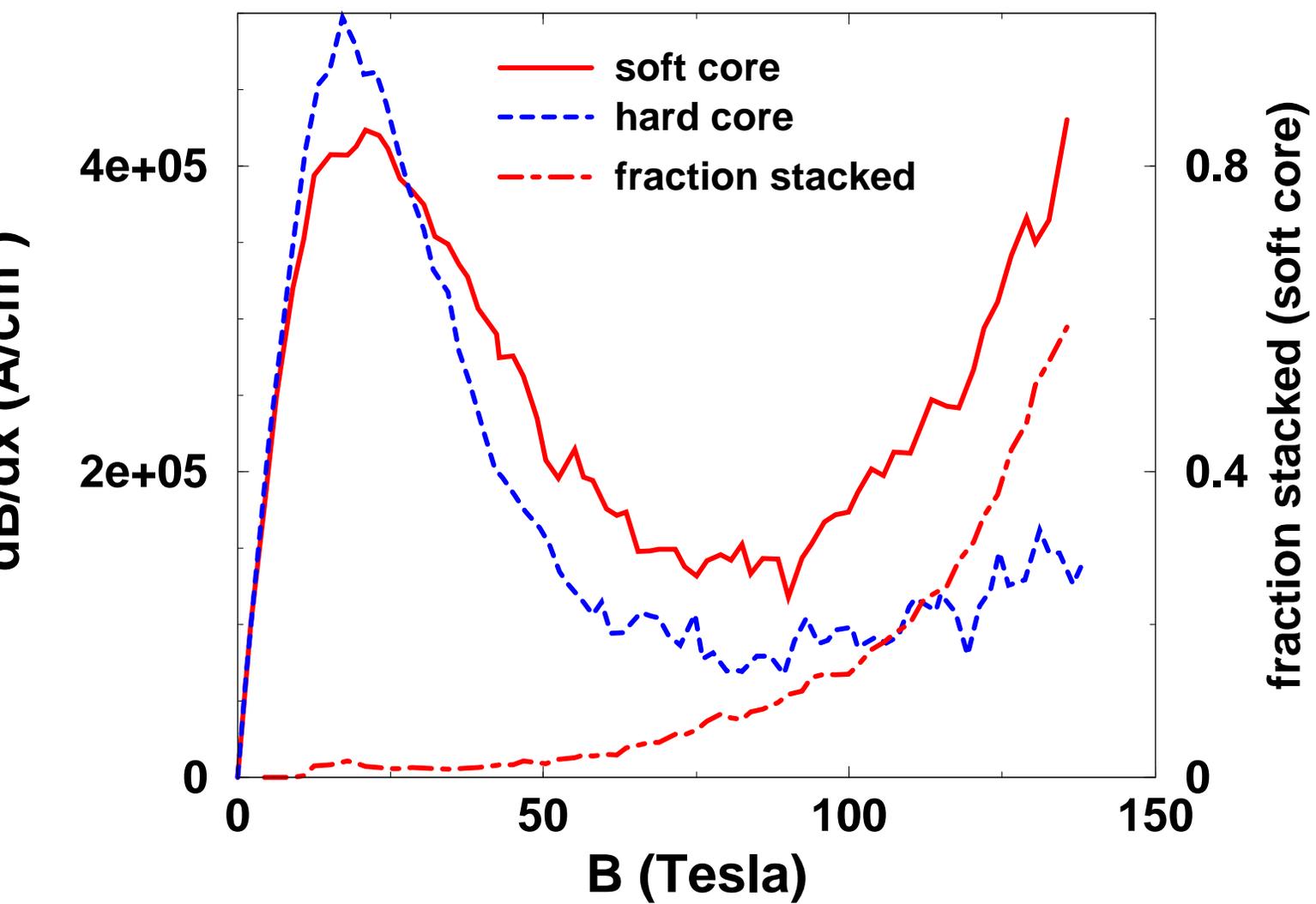
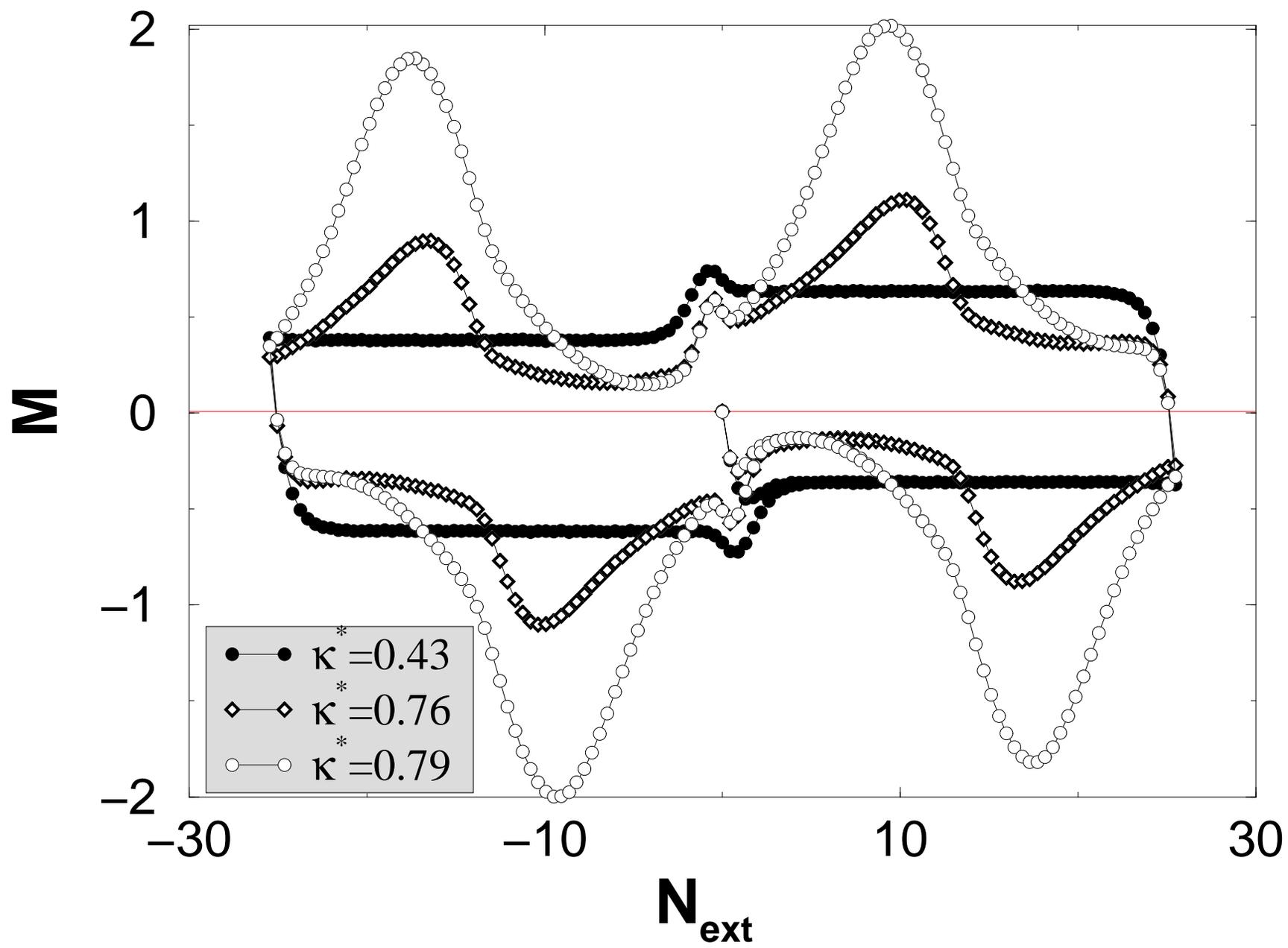
\end{document}